\documentclass[twocolumn,twocolappendix]{aastex7}

\usepackage{soul}
\newcommand{\yc}{\color{black}}

\begin{document}

\title{A Bridge between Young Stars Formed by Gravitational Interaction}

\author[0009-0004-9279-780X]{Youngwoo Choi}
\affiliation{Department of Physics and Astronomy, Seoul National University, Gwanak-ro 1, Gwanak-gu, Seoul, 08826, Korea}
\email{}

\author[0000-0003-4022-4132]{Woojin Kwon}
\affiliation{Department of Earth Science Education, Seoul National University, Gwanak-ro 1, Gwanak-gu, Seoul, 08826, Korea}

\affiliation{SNU Astronomy Research Center, Seoul National University, Gwanak-ro 1, Gwanak-gu, Seoul, 08826, Korea}

\affiliation{The Center for Educational Research, Seoul National University, 1 Gwanak-ro, Gwanak-gu, Seoul 08826, Republic of Korea}
\email{}

\author[0000-0002-4540-6587]{Leslie W. Looney}
\affiliation{Department of Astronomy, University of Illinois, 1002 West Green Street, Urbana, IL 61801, USA}
\email{}

\author[0000-0002-6195-0152]{John J. Tobin}
\affiliation{National Radio Astronomy Observatory, 520 Edgemont Road, Charlottesville, VA, 22093, USA}
\email{}

\correspondingauthor{Woojin Kwon}
\email{wkwon@snu.ac.kr}

\begin{abstract}

The majority of stars are born in clustered environments. In these environments, close encounters between young stars with planet-forming disks are expected to occur frequently. However, direct evidence of such interactions remains rare. Here, we report clear signatures of a recent dynamical interaction between the young stellar systems L1448 IRS3A and L1448 IRS3B. Millimeter wavelength observations reveal a distinct, tidally stripped bridge between the two systems. Together with previously reported spiral arm structures {\yc at the outer edge of the IRS3A disk and at the inner and outer regions of the IRS3B disk}, these features imply that the systems are undergoing a dynamical flyby. Hydrodynamical simulations reproduce these features and suggest that the closest approach occurred about 15,000 yr ago. These findings offer rare insight into how stellar interactions can reshape disk structures and influence the formation of young multiple stellar systems.

\end{abstract}


\section{Introduction} \label{sec:intro}

Most stars form in a clustered environment \citep{Lada_2003}. {\yc For instance, the stellar mass density in the NGC 1333 star-forming region is $\sim$1000 $M_{\odot}$ pc$^{-3}$ \citep{Parker_2017}, which is several orders of magnitude higher than that of the solar neighborhood, $\sim$0.04 $M_{\odot}$ pc$^{-3}$ \citep{McKee_2015}. Therefore, a} large fraction of young stars with planet-forming disks are thought to be dynamically perturbed by other stars during their protostellar stage, as they are born in highly dense surroundings \citep{Pfalzner_2013,Cuello_2023}. The dynamical interaction between two stellar systems, referred to as a flyby, occurs when a system on a parabolic or hyperbolic orbit perturbs another system \citep{Cuello_2023}. In particular, when such interactions involve young stars, their accretion disks can be strongly affected due to their large spatial extent. Therefore, {\yc flybys} can strip and truncate the planet-forming disks around young stars, affecting {\yc the subsequent evolution of their planetary systems}. 

The primary consequences of a flyby between young stars with disks include the formation of a long bridge between the interacting objects and the generation of spiral arms in their disks \citep{Clarke_1993,Cuello_2019,Kuffmeier_2019,Cuello_2020}. Under typical star-forming conditions, more than 50\% of young stars with planet-forming disks are predicted to experience a stellar flyby \citep{Cuello_2023}. However, observable signatures of such events, such as long, tidally stripped bridges, tend to dissipate on short timescales (a few thousand years for a typical disk size) \citep{Cuello_2023}, making it challenging to capture flybys in action. As a result, only a small number of flyby candidates have been identified to date \citep[e.g.,][]{Dai_2015,Winter_2018,Kurtovic_2018,Zapata_2020,Menard_2020,Dong_2022,Lu_2022}.

Depending on the angle ($\beta$) between the angular momentum vector of the perturber's orbit and that of the disk, flybys are categorized as either prograde ($-90^{\circ} < \beta < 90^{\circ}$) or retrograde ($90^{\circ} < \beta < 270^{\circ}$). In a prograde flyby, two prominent spiral arms are typically generated: the primary arm appears as a bridge between the disk and the perturber, while the secondary arm forms a single spiral structure in the disk, oriented opposite to the primary arm \citep{Cuello_2019, Cuello_2020}. In contrast, a retrograde flyby generally has a milder impact on the disk, often resulting in disk warping rather than significant stripping of material, unless the encounter is deeply penetrating \citep{Cuello_2019, Cuello_2020}.

L1448 IRS3 is located in the L1448 region of the Perseus molecular cloud at a distance of approximately 300 pc \citep{Ortiz_2018, Zucker_2018}. The system comprises three young stellar objects: L1448 IRS3A, L1448 IRS3B, and L1448 IRS3C \citep{Looney_2000}. {\yc L1448 IRS3A is classified as a Class I protostar \citep{Tobin_2007,Tobin_2016_2}, whereas L1448 IRS3B and IRS3C are Class 0 sources,} still embedded in dense, collapsing envelope material \citep{Looney_2000,Tobin_2016_2,Murillo_2016}. L1448 IRS3A and L1448 IRS3B are separated by $7^{\prime\prime}$ on the plane of the sky, corresponding to a projected distance of approximately 2100 AU \citep{Curiel_1990}. A previous study suggested that L1448 IRS3A and IRS3B are likely gravitationally bound and constitute a wide binary system, based on their projected separation and relative velocity \citep{Kwon_2006}. The third source, L1448 IRS3C, also known as L1448 NW, is located about $20^{\prime\prime}$ northwest of the other two \citep{Looney_2000}.

High-resolution ALMA observations have revealed that L1448 IRS3B is a hierarchical system, consisting of an inner binary (IRS3B-a and -b) and a more distant third component (IRS3B-c), all embedded within a common circumstellar disk \citep{Tobin_2016}. More recent studies suggest the presence of a possible fourth component, IRS3B-d, located near the system’s kinematic center \citep{Reynolds_2024, Looney_2025}. The circummultiple disk exhibits prominent one-sided spiral structures, and IRS3B-c is embedded in a dust clump located along one of the spiral arms \citep{Tobin_2016, Reynolds_2021}. These features support a disk fragmentation scenario, in which a gravitationally unstable disk fragments to form stellar components, resulting in a close multiple system. Accordingly, IRS3B-c is believed to have formed through the gravitational collapse of the unstable IRS3B disk \citep{Tobin_2016}. Follow-up ALMA observations further revealed that not only IRS3B but also {\yc IRS3A exhibits signatures of possible ring-and-gap features in its accretion disk \citep{Reynolds_2021,Reynolds_2024}.} In addition, recent NOEMA observations have identified a faint bridge-like structure between IRS3A and IRS3B \citep{Gieser_2024}.

In this Letter, we investigate the flyby between L1448 IRS3A and L1448 IRS3B. Section \ref{sec:methods} outlines the observational and simulation methods. The observational results are presented in Section \ref{sec:results}, while the outcomes of the hydrodynamical and radiative transfer simulations are described in Section \ref{sec:simulation}. Finally, we summarize our conclusions in Section \ref{sec:conclusion}.

\section{Methods} \label{sec:methods}

\subsection{Observations} \label{subsec:observations}

L1448 IRS3 was observed with ALMA Band 7 during Cycle 4 (2016.1.01520.S, PI: John Tobin). The observations include two array configurations: an extended configuration (C40-6), which provides high angular resolution, and a compact configuration (C40-3), which is more sensitive to large-scale structures. The extended configuration observations were carried out on 2016 October 1 and 4, while the compact configuration observations were conducted on 2016 December 19. The gain, bandpass, and flux calibrators used were J0336+3218, J0237+2848, and J0238+1636, respectively. The dataset includes observations of the dust continuum and the following molecular line transitions: CO ($J$ = 3 → 2), C$^{17}$O ($J$ = 3 → 2), H$^{13}$CO$^{+}$ ($J$ = 4 → 3), H$^{13}$CN ($J$ = 4 → 3)/SO$_2$ ($J$ = $13_{2,12}$ → $12_{1,11}$), and SiO ($J$ = 8 → 7), where $J$ denotes the rotational quantum number. The H$^{13}$CN line is blended with the SO$_2$ line \citep{Reynolds_2021}. The data were originally published in a previous study that revealed a prominent spiral structure in the L1448 IRS3B disk \citep{Reynolds_2021}. In this study, we use only the compact configuration (C40-3) data, which offers enhanced sensitivity to large-scale structures, despite its lower spatial resolution. This makes it better suited for investigating the larger-scale morphology of the target. {\yc As we do not find any clear signatures of flyby interactions in the jet or outflow structures traced by CO ($J$ = 3 → 2) and SiO ($J$ = 8 → 7) observations, they are not analyzed in this Letter.}

The raw data were calibrated and imaged using the Common Astronomy Software Applications (CASA) package {\yc version 4.7.0.} Phase-only self-calibration was performed on the dust continuum data with an infinite and a 60 s solution interval. The resulting phase solutions were applied to both the continuum and spectral line data to improve the signal-to-noise ratio. Continuum imaging was conducted using the CASA task \texttt{tclean} with Briggs weighting and a robust parameter of 0.5. The synthesized beam size is $0.758^{\prime\prime} \times 0.442^{\prime\prime}$ with a position angle (PA) of $-15.7^{\circ}$. The resulting continuum image, presented in Figure \ref{fig:cont}, has an rms noise level of 0.3 mJy beam$^{-1}$. Molecular line imaging was also carried out with \texttt{tclean}, using the same weighting scheme as the continuum. Integrated intensity (moment 0) and intensity-weighted velocity (moment 1) maps for the H$^{13}$CO$^{+}$ ($J$ = 4 → 3), C$^{17}$O ($J$ = 3 → 2), and H$^{13}$CN ($J$ = 4 → 3)/SO$_2$ ($J$ = $13_{2,12}$ → $12_{1,11}$) transitions are shown in Figures \ref{fig:line}. Corresponding channel maps are provided in Appendix \ref{sec:channel}. 

\subsection{Hydrodynamical Simulations} \label{subsec:simulations}

\begin{deluxetable}{lcc}
\tablecaption{Simulation setups for the L1448 IRS3A–IRS3B flyby runs \label{tab:setup}}
\tablewidth{0pt}
\tablehead{
\colhead{Parameter} & \colhead{Run\,1: Without an Embedded Source} & \colhead{Run\,2: With an Embedded Source}
}
\startdata
\multicolumn{3}{l}{\textbf{Protostars}}\\
Mass (IRS3A / IRS3B) & \multicolumn{2}{c}{$1.4\,M_\odot$ / $1.15\,M_\odot$} \\
\hline
\multicolumn{3}{l}{\textbf{Disks}}\\
Mass (IRS3A / IRS3B) & \multicolumn{2}{c}{$0.04\,M_\odot$ / $0.3\,M_\odot$\tablenotemark{a}} \\
Radius (IRS3A / IRS3B) & \multicolumn{2}{c}{$500\,\mathrm{au}$ / $700\,\mathrm{au}$} \\
Inclination (IRS3A / IRS3B) & \multicolumn{2}{c}{$70^\circ$ / $135^\circ$} \\
PA (IRS3A / IRS3B) & \multicolumn{2}{c}{$135^\circ$ / $45^\circ$} \\
\hline
\multicolumn{3}{l}{\textbf{Embedded source (Run\,2 only)}}\\
Mass & -- & $50\,M_{\mathrm{J}}$ \\
Orbital radius & -- & $300$ au \\
\hline
\multicolumn{3}{l}{\textbf{Flyby orbit }}\\
Position angle of the ascending node & \multicolumn{2}{c}{$15^\circ$} \\
Inclination & \multicolumn{2}{c}{$90^\circ$} \\
Pericenter distance $r_{\rm peri}$ & \multicolumn{2}{c}{$1500\,\mathrm{au}$} \\
Initial separation & $10\,r_{\rm peri}$ & $3\,r_{\rm peri}$\tablenotemark{b} \\
\enddata
\tablenotetext{a}{Disk self-gravity is not considered.}
\tablenotetext{b}{The Run 2 simulation adopted an initial separation of $3\,r_{\rm peri}$ to minimize the impact of the embedded source on the disk before the flyby interaction.}
\end{deluxetable}

{\yc We simulate the flyby between L1448 IRS3A and L1448 IRS3B using the smoothed particle hydrodynamics (SPH) code \texttt{PHANTOM} \citep{Price_2018}. The two disks are modeled with $5 \times 10^{5}$ gas SPH particles, and each protostar is represented by a sink particle located at the center of its respective disk. For IRS3A, we adopt a central protostellar mass of 1.4 $M_{\odot}$, based on previous kinematic analysis \citep{Reynolds_2021,Gieser_2024}. In IRS3B, the combined mass of the inner binary is 1.15 $M_{\odot}$, while the mass of the tertiary component (L1448 IRS3B-c) is not well constrained but is roughly estimated to be below 0.2 $M_{\odot}$ \citep{Reynolds_2021}. We performed two simulations with and without the embedded source. In the first simulation (Run 1), the tertiary component is not included, and IRS3B is represented by a single central source of 1.15 $M_{\odot}$. In the second simulation (Run 2), an embedded source with a mass of 50 $M_{\mathrm{J}}$ is added within the IRS3B disk at an orbital radius of 300 au. The sink radius is set to 0.5 au for all sink particles.}

{\yc The initial disk masses are set to 0.04 $M_{\odot}$ for L1448 IRS3A and 0.3 $M_{\odot}$ for L1448 IRS3B, based on previous dust continuum flux measurements assuming a gas-to-dust ratio of 100 \citep{Tobin_2016, Reynolds_2021}. Although the IRS3B disk is massive enough to be gravitationally unstable, self-gravity is not included in the simulations. Specifically, only sink–sink and gas–sink gravitational interactions are considered, while gas–gas self-gravity is neglected. The discrepancies between the observations and simulations may result from the neglect of self-gravity. Nevertheless, the observed key morphological features are still reproduced by the flyby, even without including self-gravity (see Section 4). The initial radial extents of the disks are defined as $R_{\mathrm{in}} = 10$ au and $R_{\mathrm{out}} = 500$ au for L1448 IRS3A, and $R_{\mathrm{in}} = 20$ au and $R_{\mathrm{out}} = 700$ au for L1448 IRS3B. The initial position angle and inclination of the L1448 IRS3A disk are set to 135$^{\circ}$ and 70$^{\circ}$, respectively, while those for the L1448 IRS3B disk are 45$^{\circ}$ and 135$^{\circ}$. Both disks initially follow a power-law surface density profile of $\Sigma \propto R^{-1}$, and we adopt a Shakura–Sunyaev disk viscosity parameter of $\alpha_{\mathrm{ss}} = 0.005$.}

{\yc We use the built-in flyby setup provided in the \texttt{PHANTOM} repository. In the simulations, L1448 IRS3A is modeled as the perturber on a parabolic orbit around L1448 IRS3B. The geometry of the parabolic orbit is defined by two parameters: the position angle of the ascending node and the orbital inclination. These are set to 15$^{\circ}$ and 90$^{\circ}$, respectively. The angle $\beta$, which is defined as the angle between the angular momentum vectors of the perturber’s orbit and the disk, is calculated to be approximately 52$^{\circ}$, corresponding to a prograde flyby. The pericenter distance is set to $r_{\mathrm{peri}} = 1500$ au, and the initial separation between the two stars is 10 times the pericenter distance. However, the simulation with the embedded source adopted an initial separation of 3 times the pericenter distance, as the embedded source significantly affects the disk mass and morphology even before the flyby. To reproduce the observed morphology, we apply a 180$^{\circ}$ rotation of the system around the $z$-axis, corresponding to the line-of-sight direction. The simulation setups are summarized in Table \ref{tab:setup}.}

\section{Results} \label{sec:results}

\subsection{Dust Continuum} \label{subsec:continuum}

\begin{figure*}[ht!]
\epsscale{1}
\plotone{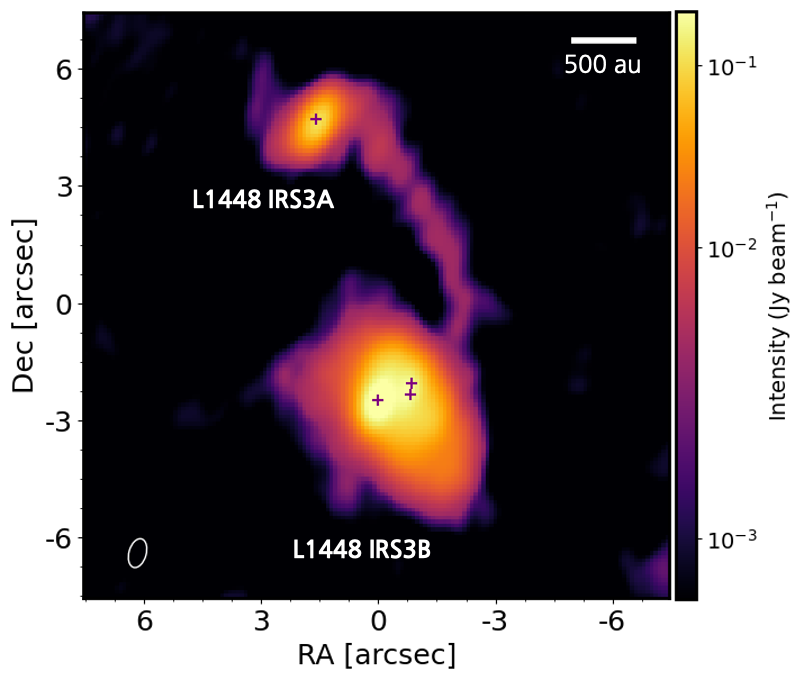}
\caption{ALMA 870 $\mu$m dust continuum image of the L1448 IRS3A and L1448 IRS3B systems. The upper source is L1448 IRS3A, and the lower is L1448 IRS3B. The image was produced using Briggs weighting with a robust parameter of 0.5. The synthesized beam size is $0.758^{\prime\prime} \times 0.442^{\prime\prime}$ with a position angle of $-15.7^{\circ}$. The beam and the spatial scale bar are shown in the lower left and upper right corners of the panel, respectively. The positions of the protostars are indicated by plus symbols. {\yc The central coordinates (ICRS) are R.A. = 03$^{\mathrm{h}}$25$^{\mathrm{m}}$36.385$^{\mathrm{s}}$, decl. = +30$^{\circ}$45$^{\prime}$17.2$^{\prime\prime}$.}
\label{fig:cont}}
\end{figure*}

Figure \ref{fig:cont} shows the ALMA 870 $\mu$m dust continuum image toward L1448 IRS3A and L1448 IRS3B. The cross markers indicate the positions of the protostars in each system. Strikingly, L1448 IRS3A and L1448 IRS3B are connected by a long, narrow bridge-like dust structure. The length of the bridge is approximately $6^{\prime\prime}$ (projected length of 1800 AU), and the width is about $0.8^{\prime\prime}$ (projected width of 240 AU). This narrow bridge between the two young stars suggests that the systems are dynamically interacting with each other. Previous simulations have shown that such a bridge between young stellar objects is likely produced by a close encounter between a disk-hosting protostar and a perturber, which can tidally strip material from the planet-forming disk \citep{Cuello_2019, Cuello_2020}. Therefore, it is likely that L1448 IRS3A and L1448 IRS3B experienced a recent close encounter.

Previous high-resolution observations have resolved a prominent spiral arm in the L1448 IRS3B accretion disk, oriented opposite to the bridge structure \citep{Tobin_2016, Reynolds_2021,Looney_2025}. This spatial relationship suggests that the bridge and the spiral arm likely share a common origin and together form a large-scale two-arm spiral pattern. In addition to the bridge, an S-shaped, large-scale spiral structure is also observed in L1448 IRS3A. These features resemble those produced in simulations of a prograde flyby in which the perturber captures material from the disk \citep{Cuello_2023}. Taken together, these characteristics indicate that L1448 IRS3A likely perturbed the L1448 IRS3B disk through a prograde flyby. This event is responsible for the formation of (1) the bridge extending from IRS3B to IRS3A \citep{Gieser_2024},
(2) the prominent spiral arm in the IRS3B disk \citep{Tobin_2016,Reynolds_2021}, and 
(3) the S-shaped large-scale spiral in IRS3A \citep{Gieser_2024,Looney_2025}.

\subsection{Molecular Lines} \label{subsec:lines}

\begin{figure*}[ht!]
\epsscale{1.17}
\plotone{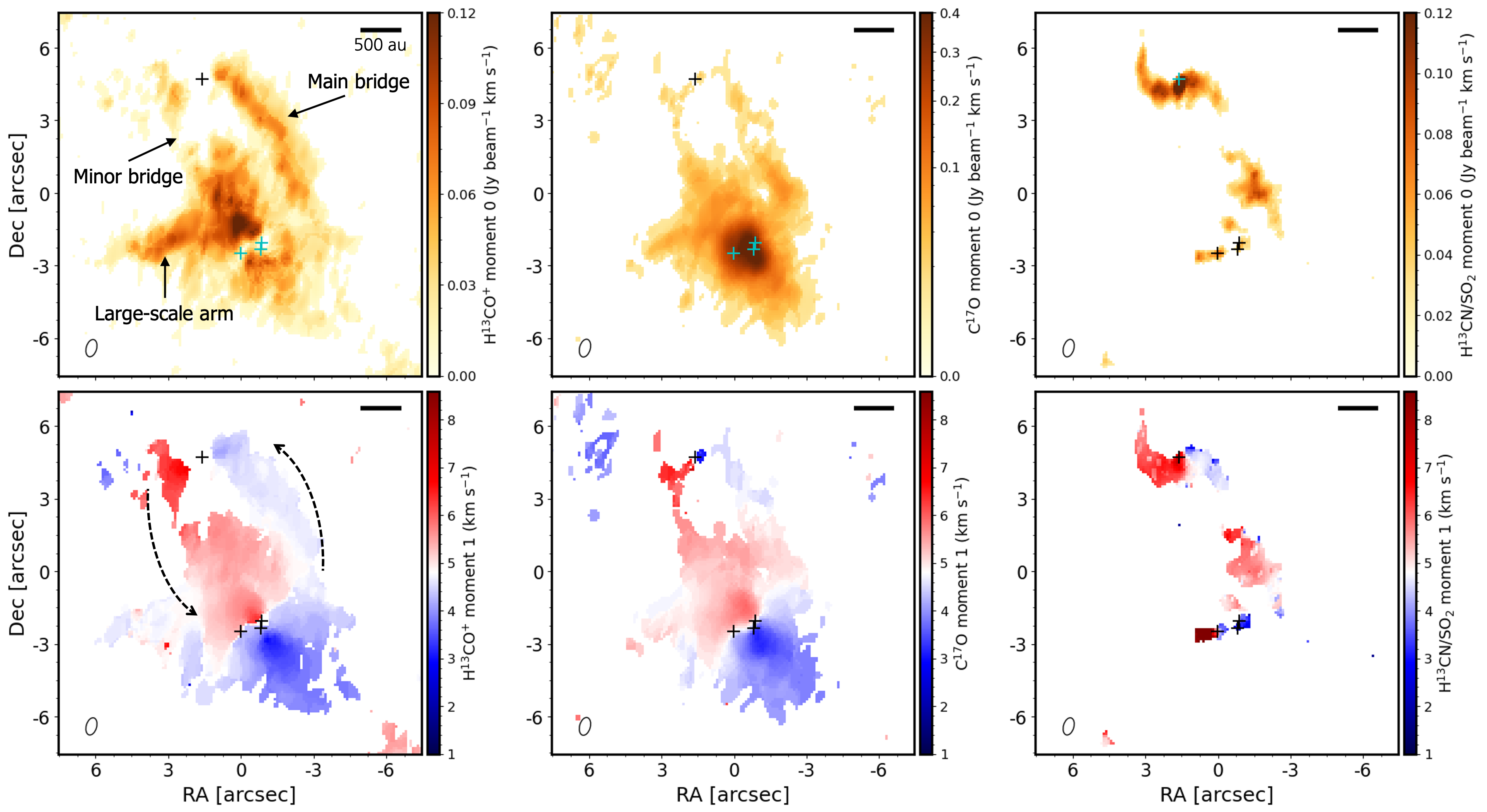}
\caption{Integrated intensity (moment 0) and intensity-weighted velocity (moment 1) maps of the H$^{13}$CO$^{+}$ (4–3) (Left), C$^{17}$O (3–2) (Middle), and H$^{13}$CN (4–3)/SO$_{2}$ ($13_{2,12}-12_{1,11}$) (Right) transitions. The synthesized beam sizes are $0.730^{\prime\prime} \times 0.430^{\prime\prime}$ with a position angle of $-15.9^{\circ}$ for H$^{13}$CO$^{+}$ (4–3), $0.753^{\prime\prime} \times 0.443^{\prime\prime}$ with a position angle of $-15.3^{\circ}$ for C$^{17}$O (3–2), and $0.742^{\prime\prime} \times 0.431^{\prime\prime}$ with a position angle of $-15.6^{\circ}$ for H$^{13}$CN (4–3)/SO$_{2}$ ($13_{2,12}-12_{1,11}$). The beam and spatial scale bar are shown in the lower left and upper right corners of each panel, respectively. The positions of the protostars are indicated by plus symbols.
\label{fig:line}}
\end{figure*}

Features associated with dynamical interaction are also evident in the molecular line data. The left and middle panels of Figure 2 present the integrated intensity (moment 0) and intensity-weighted velocity (moment 1) maps of the H$^{13}$CO$^{+}$ (4–3) and C$^{17}$O (3–2) transitions. The bridge structure seen in the dust continuum (hereafter referred to as the main bridge) is also prominent in both H$^{13}$CO$^{+}$ and C$^{17}$O emission. Interestingly, a fainter structure (hereafter the minor bridge) appears to connect to the eastern (redshifted) side of L1448 IRS3A, in contrast to the main bridge in the western (blueshifted) side. The main bridge is blueshifted with respect to L1448 IRS3B, whereas the minor bridge is redshifted, indicating that the two structures extend in opposite directions. Since L1448 IRS3A is thought to be stripping material from the disk around L1448 IRS3B, the main bridge likely traces gas flowing from IRS3B toward IRS3A. In contrast, the redshifted minor bridge appears to be directed toward the IRS3B disk. This suggests that L1448 IRS3A is located in the foreground relative to L1448 IRS3B along the line of sight. {\yc This interpretation is supported by the fact that IRS3A is brighter in the near-infrared \citep{Tobin_2007}.} In addition to the two bridge features, the H$^{13}$CO$^{+}$ and C$^{17}$O emission maps also reveal a large-scale arm extending eastward from the L1448 IRS3B disk. This structure is also seen in recent ALMA Band 4 dust continuum observations (Figure 11; \citealt{Looney_2025}).

The right panels of Figure 2 present the integrated intensity (moment 0) and intensity-weighted velocity (moment 1) maps of the H$^{13}$CN (4–3)/SO$_{2}$ ($13_{2,12}-12_{1,11}$) emission. The H$^{13}$CN/SO$_{2}$ emission is concentrated around L1448 IRS3A and is not prominent in the L1448 IRS3B disk. In IRS3A, this emission also traces the large-scale spiral structure seen in the dust continuum image. Since SO$_{2}$ emission primarily traces highly energetic shocked gas \citep{van_Gelder_2021}, {\yc if the observed emission arises from SO$_{2}$, the detection of H$^{13}$CN/SO$_{2}$ emission around L1448 IRS3A strongly suggests that IRS3A is being impacted by material stripped from the disk of L1448 IRS3B.} In the L1448 IRS3B disk, the H$^{13}$CN/SO$_{2}$ emission is associated with the inner outflow launching regions and with the boundary between the main bridge and the disk, where disk material is thought to undergo energetic perturbations.

\section{Hydrodynamical Simulation} \label{sec:simulation}

\begin{figure*}[ht!]
\epsscale{1.1}
\plotone{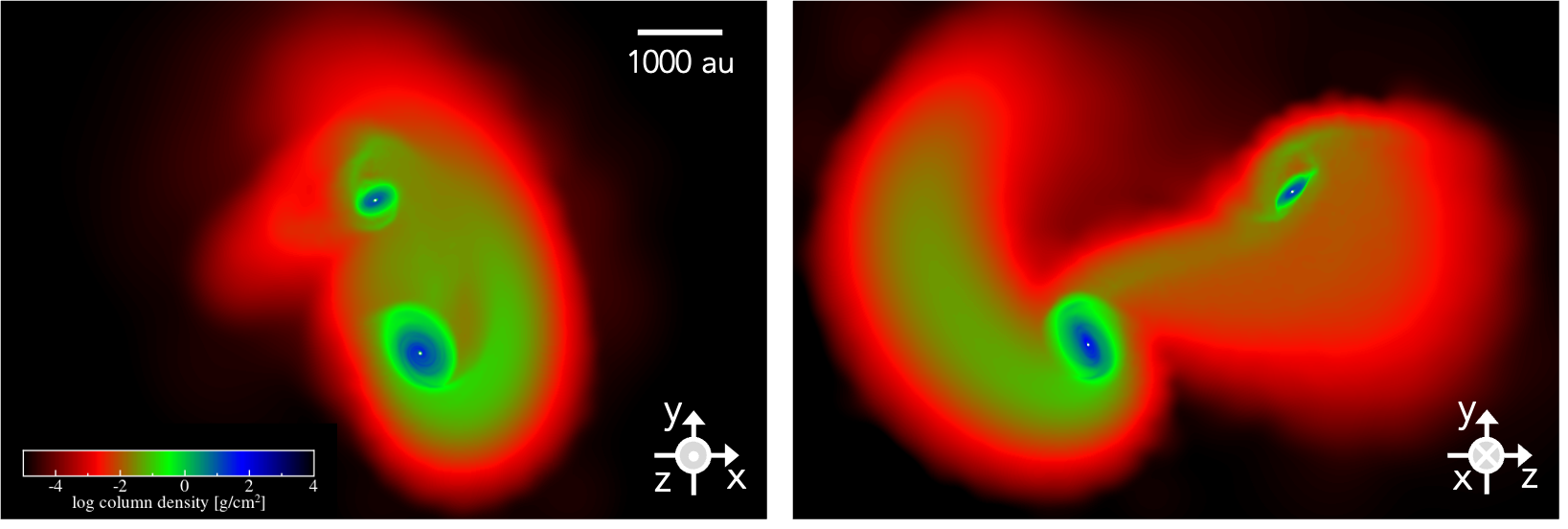}
\caption{{\yc Snapshot from the Run 1 simulation, taken 15,736 yr after the closest approach between the primary star (representing L1448 IRS3B) and the perturber (representing L1448 IRS3A). Left: viewed in the plane of the sky ($x$–$y$ plane). Right: viewed from the eastern side ($y$–$z$ plane). The color scale indicates gas column density. An animation of the simulation is available in the online Journal. The animation shows the system’s evolution over 0.15 Myr, where $t=0$ corresponds to an initial separation of 10 times the pericenter distance and the closest approach occurs at $t\sim0.094$ Myr.} 
\label{fig:video}}
\end{figure*}

We performed hydrodynamical simulations to model the flyby between L1448 IRS3A and L1448 IRS3B. {\yc Animations of Figure \ref{fig:video} show the simulated flyby between L1448 IRS3A and L1448 IRS3B (Run 1), viewed in the plane of the sky ($x$–$y$ plane) and from the eastern side ($y$–$z$ plane), respectively. Initially, L1448 IRS3A is located farther from the observer than L1448 IRS3B. In the plane-of-sky view, IRS3A appears above IRS3B, while in the side view, IRS3A is initially on the left and IRS3B on the right. Due to their mutual gravitational attraction, IRS3A moves toward IRS3B and, following the closest approach ($t=0.094$ Myr), is deflected northward (see the animations of Figure \ref{fig:video}). Figure \ref{fig:video} presents the simulation snapshot 15,736 yr after the closest approach ($t=0.11$ Myr).} The prograde flyby strongly perturbs and strips material from the L1448 IRS3B disk, with a portion of the disk material subsequently captured by L1448 IRS3A. As a result, a bridge forms between the two systems, and a spiral arm develops from the IRS3B disk in the direction opposite to the bridge.

\begin{figure*}[ht!]
\epsscale{1.1}
\plotone{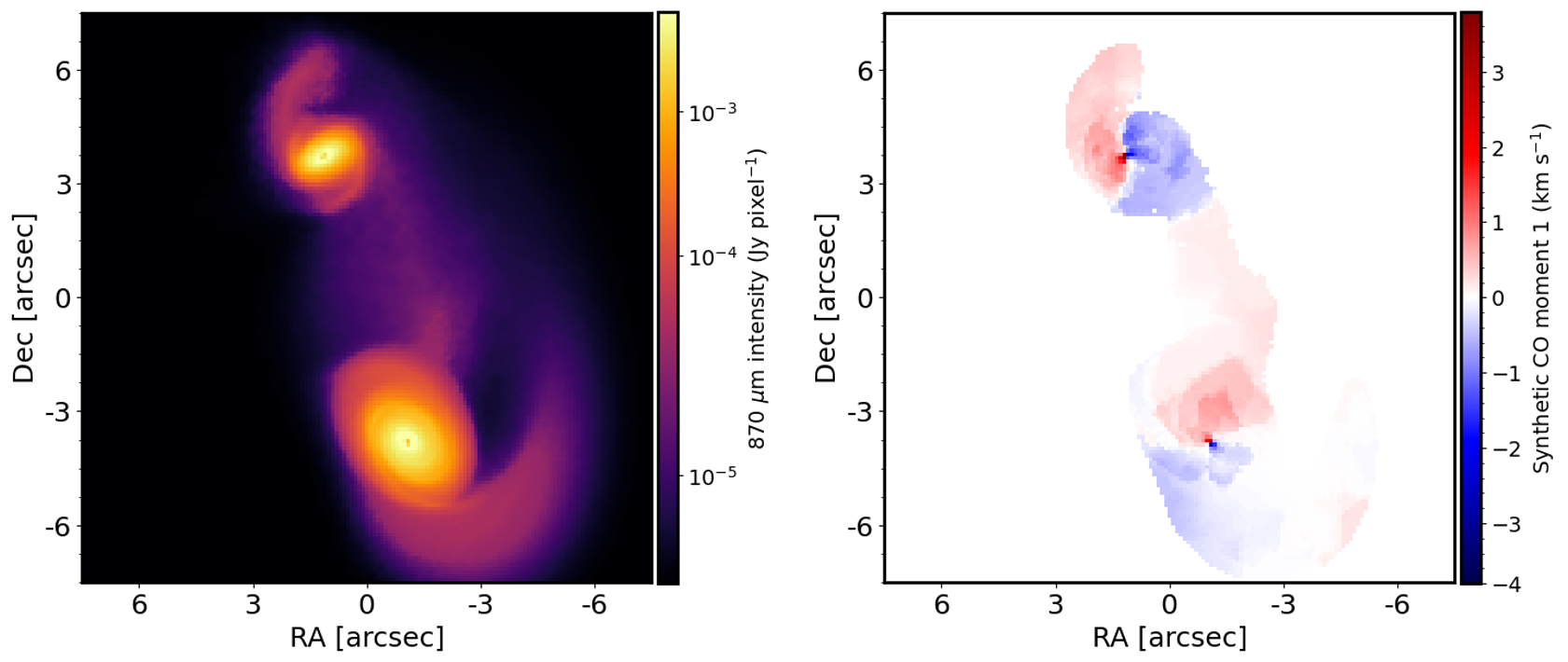}
\caption{Simulated 870 $\mu$m dust continuum and intensity-weighted velocity (moment 1) maps of CO emission {\yc for Run 1}. The hydrodynamical simulation snapshot corresponds to {\yc 15,736} yr after the perturber (representing L1448 IRS3A) passed its closest approach to the primary star (representing L1448 IRS3B). The continuum and CO velocity maps are produced using radiative transfer simulations. The CO intensity-weighted velocity map is shown only in regions where the dust continuum exceeds {\yc $1.2 \times 10^{-5}$ Jy pixel$^{-1}$. The image pixel size is set to 0.1$^{\prime \prime}$. }
\label{fig:simulation}}
\end{figure*}

We use the radiative transfer code \texttt{MCFOST} to generate synthetic observational images from the simulation snapshots \citep{Pinte_2006,Pinte_2009,Tessore_2021}. First, the temperature of the disk material is computed by heating from the two central sink particles. The effective temperatures of the stellar components are estimated using 1 Myr isochrones. The resulting effective temperatures are 4498 K for IRS3A and 4367 K for IRS3B \citep{Siess_2000}. To compute the dust temperature structure, we launch $10^{8}$ photons. The dust grain population follows a standard MRN distribution ($\mathrm{d}n(a) \propto a^{-3.5}$) with grain sizes ranging from $a_{\mathrm{min}} = 0.03$ $\mu$m to $a_{\mathrm{max}} = 1$ mm. A synthetic 870 $\mu$m dust continuum image is generated using a ray-tracing technique and is shown in the left panel of Figure \ref{fig:simulation}. The image corresponds to the simulation snapshot {\yc 15,736} yr after the closest approach. An intensity-weighted velocity (moment 1) map is also created from the synthetic CO emission cube using the same snapshot. The resulting moment 1 map is presented in the right panel of Figure \ref{fig:simulation}.

The synthetic dust continuum image successfully reproduces three key observational features: the bridge between the two systems, the spiral arm in the L1448 IRS3B disk, and the large-scale spiral structure in L1448 IRS3A. It also predicts a prominent spiral arm extending westward from the southern region of the IRS3B disk, a feature that is partially seen in recent dust continuum observations (Figure 11; \citealt{Looney_2025}). The synthetic line-of-sight velocity structure is also broadly consistent with the observations. The main bridge appears redshifted in the IRS3B disk and gradually becomes blueshifted toward the western side of L1448 IRS3A. 

However, some discrepancies remain between the observations and the simulation. The redshifted bridge connected to the eastern side of L1448 IRS3A (referred to as the minor bridge) is not reproduced in the simulation. In addition, the large-scale arm extending eastward in the observations is not clearly seen in the synthetic images. {\yc Similar discrepancies between the observations and the simulation are also found in other flyby candidates. For instance, although the spiral morphology of the UX Tau system is generally well reproduced by flyby simulations, the outermost spiral feature cannot be explained \citep{Menard_2020}. Likewise, while the flyby simulation can reproduce the main tidal arm in the RW Aur system \citep{Cabrit_2006,Dai_2015}, the other observed tidal streams cannot be explained by a single star–disk tidal encounter \citep{Rodriguez_2018}.}

One possible explanation for {\yc the discrepancies in the L1448 IRS3 system} is that the simulation does not account for the dense ambient material surrounding L1448 IRS3A and IRS3B. {\yc Both the IRS3A and IRS3B disks are embedded within dense envelope material \citep{Terebey_1997, Looney_2000}.} Recent ALMA observations have also revealed various asymmetric structures in the protostellar envelope, such as streamer-like features \citep{Gieser_2024,Gieser_2025}, suggesting that the actual flyby between IRS3A and IRS3B may involve such envelope structures. In contrast, our simulation includes only the circumstellar disks of the two systems. {\yc Another possible explanation is the disk’s self-gravity. As gravitational instability can generate spiral arms in disks \citep[e.g.,][]{Kratter_2016}, the combined effects of the flyby and gravitational instability might produce the observed large-scale arm that is not reproduced in the current simulation. This possibility should be tested with simulations that include self-gravity.} 

\begin{figure*}[ht!]
\epsscale{1}
\plotone{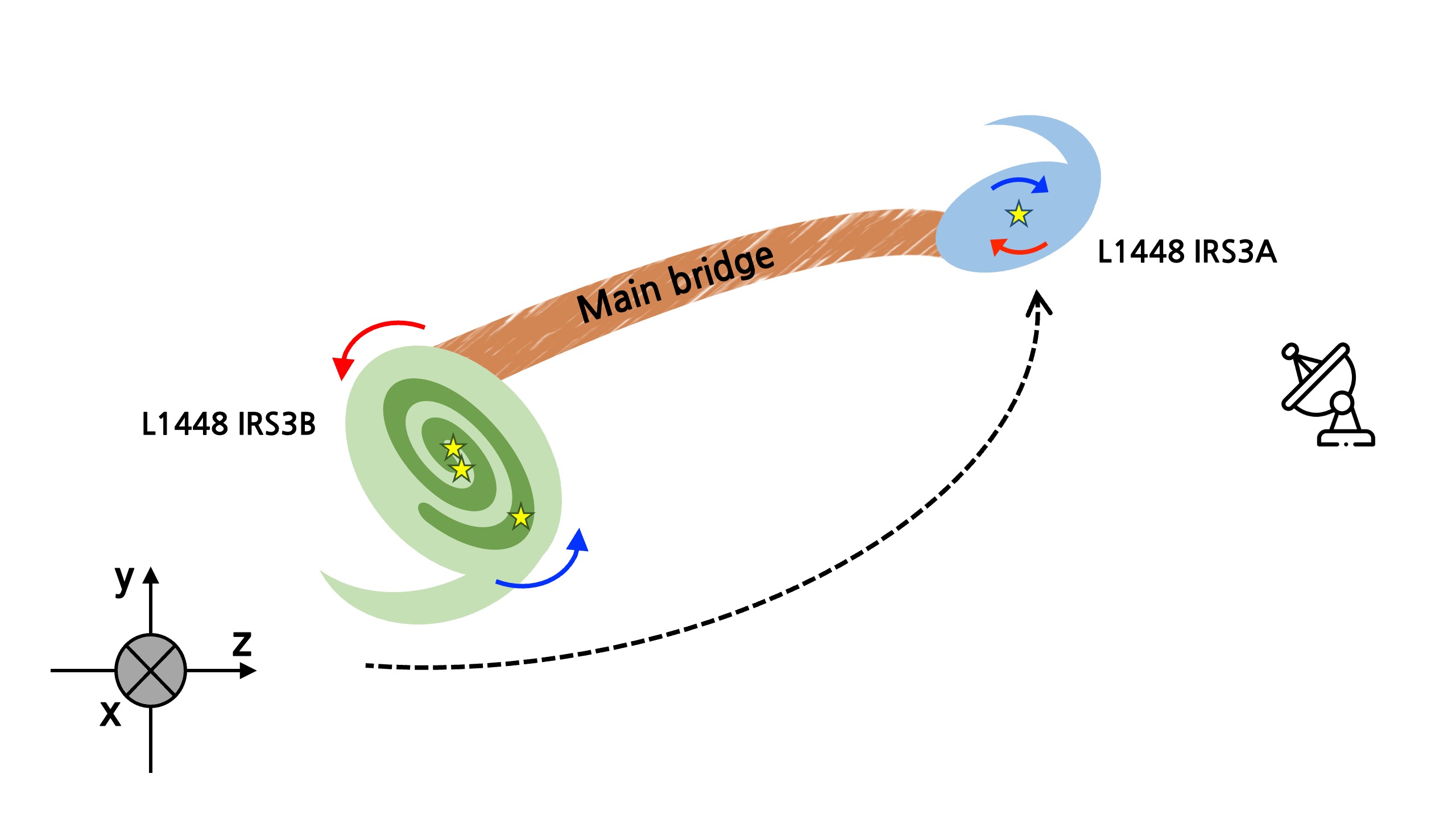}
\caption{Schematic illustration of the flyby simulation geometry. The z-axis corresponds to the line-of-sight direction. The main bridge is shown in orange. The disk and spiral structure of L1448 IRS3A are shown in blue, while those of L1448 IRS3B are shown in green.
\label{fig:cartoon}}
\end{figure*}

Moreover, while kinematic analysis shows that L1448 IRS3A is redshifted by approximately 0.5 km s$^{-1}$ relative to IRS3B \citep{Reynolds_2021}, the simulation yields almost no line-of-sight velocity difference. This discrepancy may arise from slight variations in the flyby orbit or the viewing geometry. Nevertheless, the main bridge and spiral structures in both systems are clearly reproduced by the simulation. The schematic geometry of the flyby simulation is illustrated in Figure \ref{fig:cartoon}.

The spiral structures observed in the L1448 IRS3B disk might have existed before the flyby with L1448 IRS3A, as the disk is sufficiently massive to become gravitationally unstable \citep{Tobin_2016}. Nevertheless, the flyby may have played a significant role in shaping the disk, as the spirals extend in the direction opposite to the bridge \citep{Looney_2025}, consistent with a flyby scenario. Since L1448 IRS3B-c, the tertiary component in L1448 IRS3B, is located within the spiral structure of the IRS3B disk, its {\yc evolution} is likely linked to the disk substructure. {\yc The protostellar mass of L1448 IRS3B-c is estimated to be below 0.2 $M_{\odot}$, inferred from the lack of noticeable perturbations in the surrounding disk kinematics \citep{Reynolds_2021}. To investigate how the flyby influences the evolution of L1448 IRS3B-c and how the presence of IRS3B-c affects the flyby outcome, we performed another flyby simulation that includes an embedded source (Run 2). We added an embedded source (representing L1448 IRS3B-c) to the primary disk (representing the L1448 IRS3B disk) with an initial mass of 50 Jupiter masses ($\sim$ 0.05 $M_{\odot}$) and an initial orbital radius of 300 au. The mass and orbital radius evolve to $\sim$0.1 $M_{\odot}$ and 250 au at the estimated current epoch. The other simulation setups are the same as the simulation without the embedded source (Run 1), except that the initial separation between the two stars is set to 3 times the pericenter distance, since the presence of the embedded source significantly influences the IRS3B disk prior to the flyby interaction.}

\begin{figure*}[ht!]
\epsscale{1.1}
\plotone{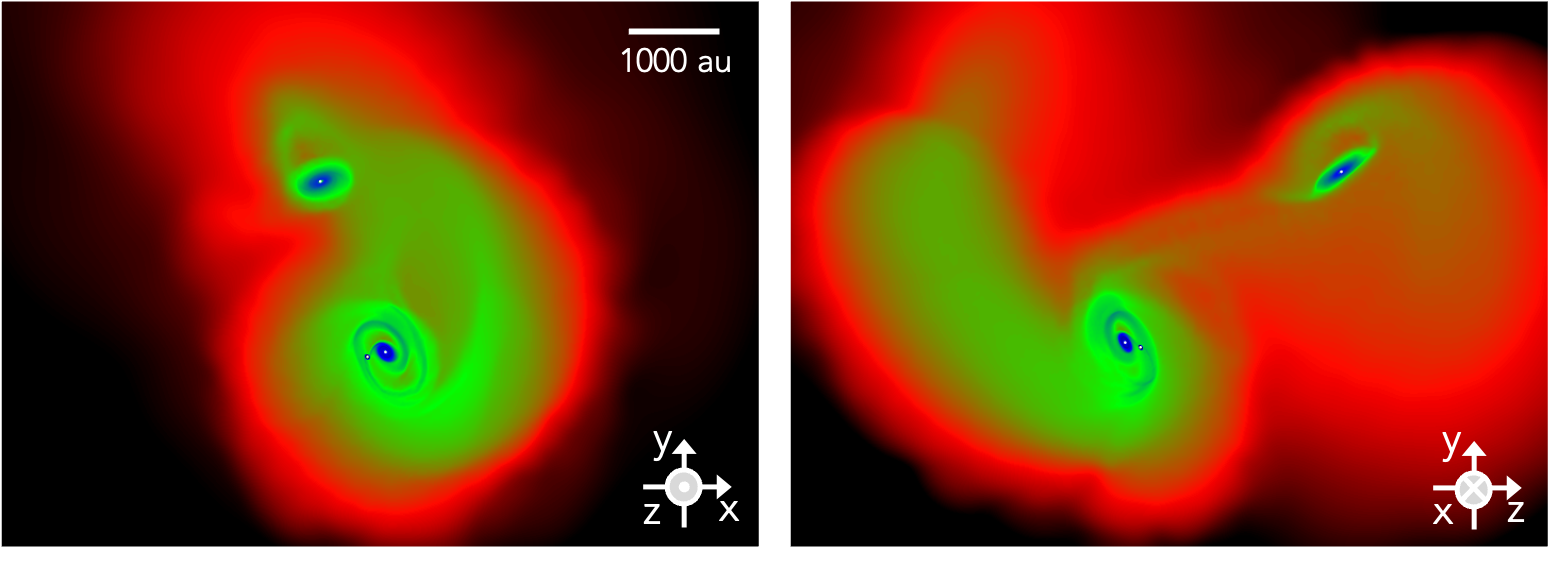}
\caption{{\yc Snapshot from the Run 2 simulation, taken 13,122 yr after the closest approach between the primary star (representing L1448 IRS3B) and the perturber (representing L1448 IRS3A). Left: viewed in the plane of the sky ($x$–$y$ plane). Right: viewed from the eastern side ($y$–$z$ plane). The color scale shows the gas column density, the same as in Figure \ref{fig:video}. An animation of the simulation is available in the online Journal. The animation shows the system’s evolution over 0.044 Myr, where $t=0$ corresponds to an initial separation of 3 times the pericenter distance and the closest approach occurs at $t\sim0.021$ Myr.} 
\label{fig:video2}}
\end{figure*}

\begin{figure*}[ht!]
\epsscale{1.1}
\plotone{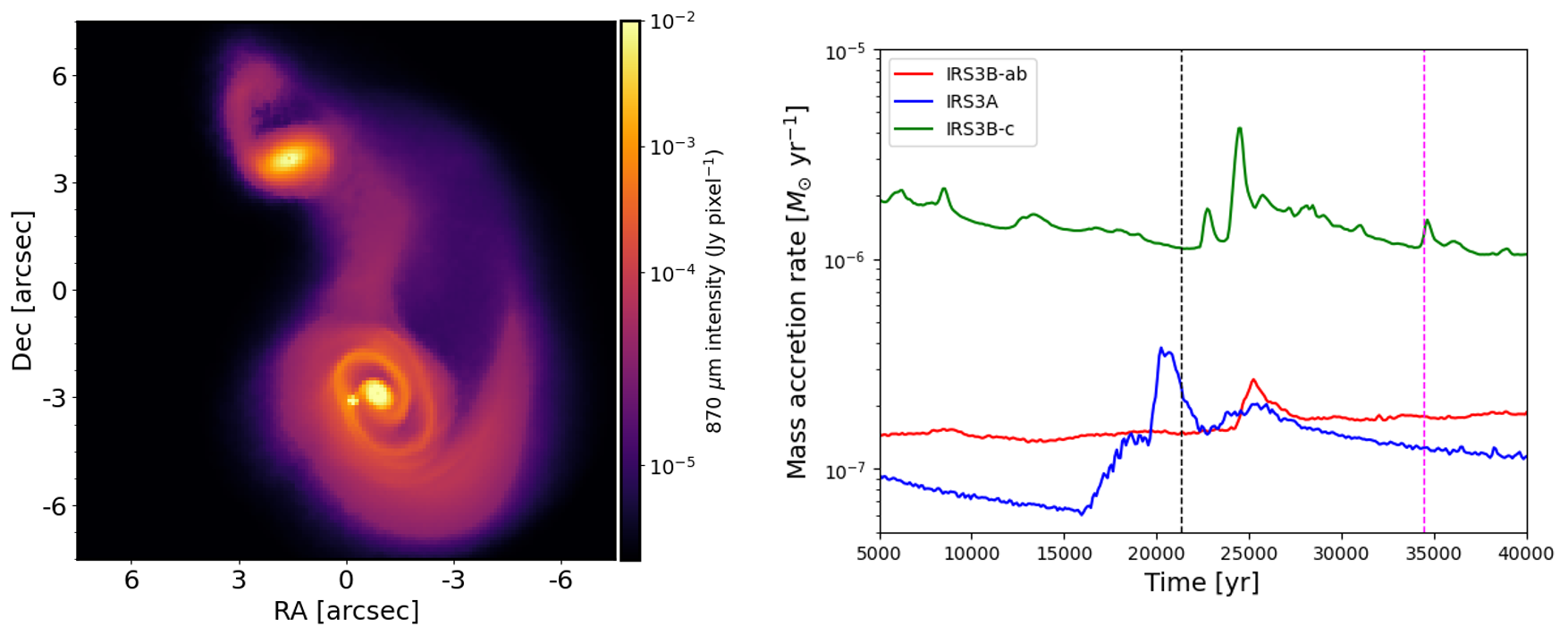}
\caption{{\yc Left: simulated 870 $\mu$m dust continuum image for Run 2. The hydrodynamical simulation snapshot corresponds to 13,122 yr after the perturber (representing L1448 IRS3A) passed its closest approach to the primary star (representing L1448 IRS3B). Right: mass accretion rates of the primary (IRS3B-ab), the perturber (IRS3A), and the embedded source (IRS3B-c) during the Run 2 simulation. The black and magenta vertical lines indicate the times of closest approach and of the snapshot shown in the left panel, respectively.} 
\label{fig:simulation2}}
\end{figure*}

{\yc Animations in Figure \ref{fig:video2} show the simulated flyby including an embedded source (Run 2). Similar to the simulation without the embedded source (Run 1), the Run 2 simulation reproduces the bridge and large-scale spiral structures. In addition, it generates distinctive spiral substructures within the IRS3B disk. These spirals are primarily driven by the embedded source and evolve into unique patterns following the flyby. The left panel of Figure \ref{fig:simulation2} presents the simulated 870 $\mu$m dust continuum image from Run 2, revealing a characteristic spiral pattern. This morphology closely resembles the spiral substructures observed in the IRS3B disk by previous high-resolution studies \citep{Tobin_2016, Reynolds_2021}. Thus, the spiral patterns can be well explained by a flyby interaction involving an embedded source.}

{\yc To examine how the flyby affects the evolution of L1448 IRS3B-c, we calculated the mass accretion rates of the system components. The right panel of Figure \ref{fig:simulation2} shows the accretion rates of the primary star (representing L1448 IRS3B-ab), the perturber (representing L1448 IRS3A), and the embedded source (representing L1448 IRS3B-c) during the flyby simulation. The accretion rate of the perturber increases near the time of closest approach as it captures material from the primary disk. After the encounter, the mass accretion rates of both the primary and the embedded source rise significantly, indicating that the flyby triggers enhanced accretion onto these components \citep{Cuello_2019}. The accretion rate of the embedded source remains higher than those of the primary and the perturber throughout the subsequent evolution. This high mass accretion rate may explain the bright compact structure observed around IRS3B-c in the L1448 IRS3B disk \citep[e.g.,][]{Tobin_2016}. We note that the precise accretion rate depends on the assumed mass of IRS3B-c, which should be better constrained by future high-resolution observations.}

\section{Conclusions} \label{sec:conclusion}
We present ALMA Band 7 observations of the L1448 IRS3A and L1448 IRS3B protostellar systems. The dust continuum data reveal a prominent bridge connecting the two systems. L1448 IRS3A exhibits large-scale spiral structures, while previous high-resolution studies have identified a one-sided spiral arm in the L1448 IRS3B disk extending opposite to the bridge \citep{Tobin_2016,Reynolds_2021}. Such a morphology is consistent with a recent prograde flyby between the two systems.

The H$^{13}$CO$^{+}$ ($J$ = 4 → 3) and C$^{17}$O ($J$ = 3 → 2) observations not only trace the main bridge seen in the dust continuum but also reveal a fainter secondary bridge and eastward arm-like structures. Emission from H$^{13}$CN ($J$ = 4 → 3)/SO$_{2}$ ($J$ = $13_{2,12}$ → $12_{1,11}$) is concentrated toward L1448 IRS3A. {\yc If the H$^{13}$CN/SO$_{2}$ emission mainly originates from SO$_{2}$, it indicates that the L1448 IRS3A system is undergoing strong energetic perturbations.}

We performed flyby simulations of the L1448 IRS3A–IRS3B interaction and found that the dynamical encounter can reproduce the prominent bridge, the one-sided spiral substructures in the IRS3B disk, and the large-scale spirals in the IRS3A disk. However, the secondary bridge and eastward arm-like structures are not reproduced, likely since our simulations do not account for the surrounding protostellar envelope {\yc and disk self-gravity}. The hydrodynamical models suggest that the closest approach occurred about 15,000 yr ago. {\yc We also performed an additional simulation that includes an embedded source within the L1448 IRS3B disk to investigate how the flyby affects the L1448 IRS3B-c component and how the presence of IRS3B-c influences the flyby outcome. The simulation shows that a close encounter involving an embedded source can produce the distinctive spiral substructures observed in the IRS3B disk \citep{Tobin_2016, Reynolds_2021}. In addition, the IRS3B-c component maintains a high accretion rate throughout the simulation, potentially explaining the bright compact structure surrounding the IRS3B-c protostar.}

\vspace{5mm}

We are grateful to the anonymous referee for helpful comments. This work was supported by the National Research Foundation of Korea (NRF) grant funded by the Korea government (MSIT) (RS-2024-00342488 and RS-2024-00416859). This paper makes use of the following ALMA data: ADS/JAO.ALMA\#2016.1.01520.S. ALMA is a partnership of ESO (representing its member states), NSF (USA) and NINS (Japan), together with NRC (Canada), MOST and ASIAA (Taiwan), and KASI (Republic of Korea), in cooperation with the Republic of Chile. The Joint ALMA Observatory is operated by ESO, AUI/NRAO and NAOJ. The National Radio Astronomy Observatory and Green Bank Observatory are facilities of the U.S. National Science Foundation operated under cooperative agreement by Associated Universities, Inc. 

\facilities{ALMA}
\software{CASA \citep{CASA_2022}, NumPy \citep{Harris_2020}, Matplotlib \citep{Hunter_2007}, Astropy \citep{Astropy_2013,Astropy_2018,Astropy_2022}, SciPy \citep{Virtanen_2020}}

\appendix

\section{Channel maps}
\label{sec:channel}

We present channel maps of H$^{13}$CO$^{+}$ ($J$ = 4 → 3), C$^{17}$O ($J$ = 3 → 2), and H$^{13}$CN ($J$ = 4 → 3)/SO$_2$ ($J$ = $13_{2,12}$ → $12_{1,11}$). The H$^{13}$CO$^{+}$ and C$^{17}$O maps reveal a characteristic “butterfly pattern” tracing rotational motion around L1448 IRS3B in the velocity range 3.0–4.3 km s$^{-1}$. The main bridge is most clearly detected between 4.1–5.1 km s$^{-1}$, while the minor bridge appears between 5.1–6.4 km s$^{-1}$.

\begin{figure*}[ht!]
\epsscale{1.15}
\plotone{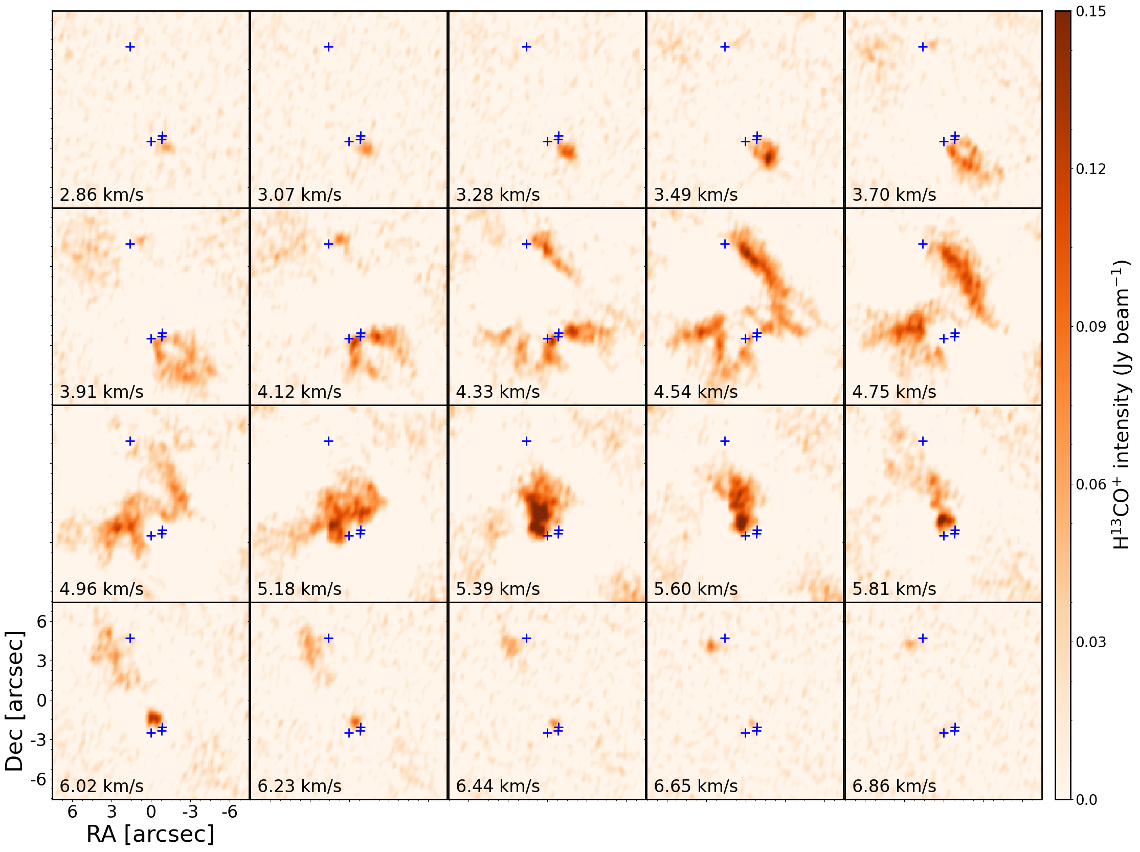}
\caption{Channel maps of the H$^{13}$CO$^{+}$ ($J$ = 4 → 3) transition. The positions of the protostars are represented as the plus indicators. The channel velocities are written on the lower left side of each map.
\label{fig:fig5}}
\end{figure*}

\begin{figure*}[ht!]
\epsscale{1.15}
\plotone{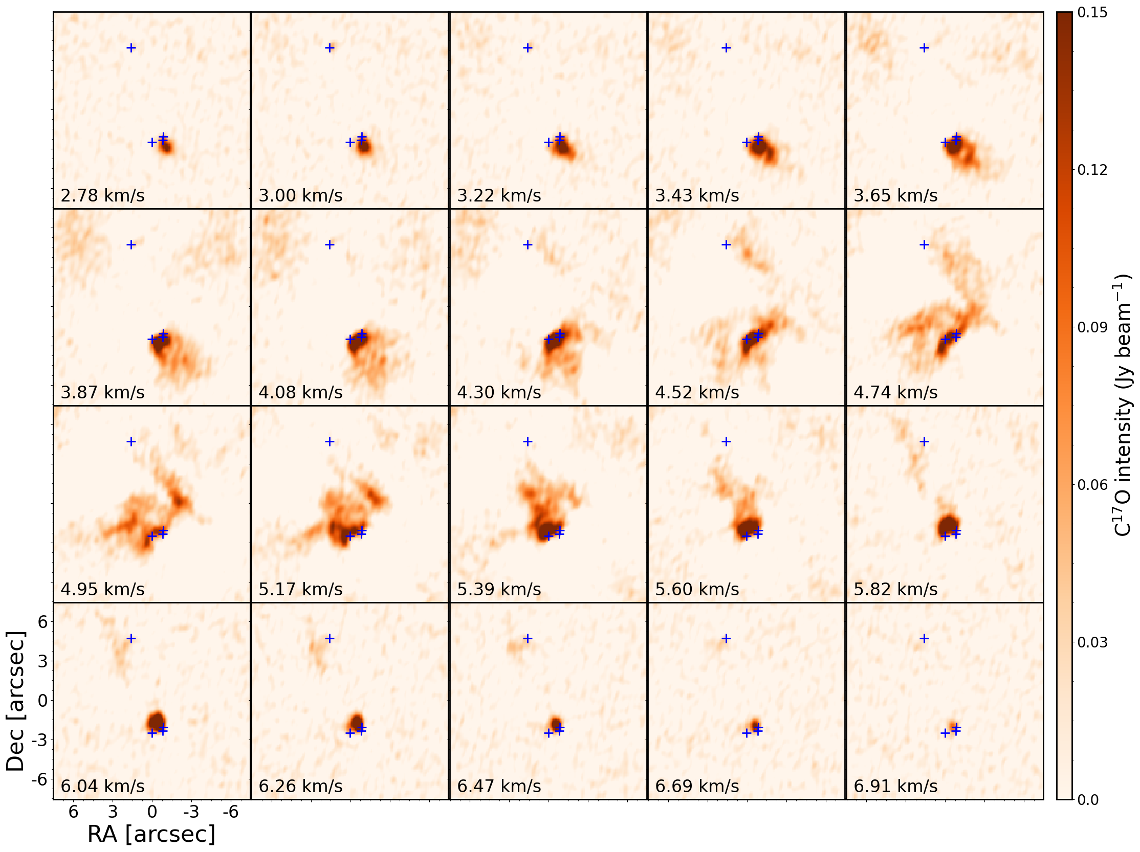}
\caption{Channel maps of the C$^{17}$O ($J$ = 3 → 2) transition. Figure \ref{fig:fig5} but for C$^{17}$O.
\label{fig:fig6}}
\end{figure*}

\begin{figure*}[ht!]
\epsscale{1.15}
\plotone{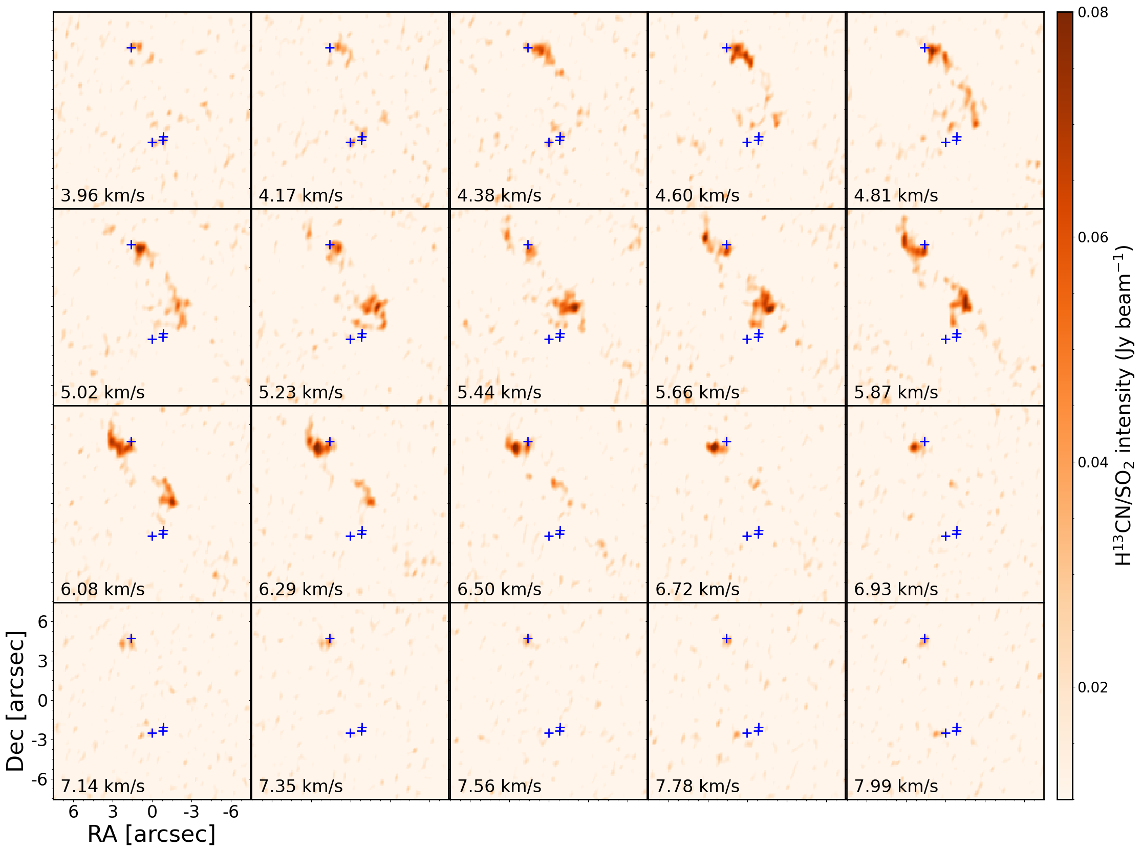}
\caption{Channel maps of the H$^{13}$CN ($J$ = 4 → 3) / SO$_{2}$ ($J$ = $13_{2,12}$ → $12_{1,11}$) transitions. Figure \ref{fig:fig5} but for H$^{13}$CN/SO$_{2}$.
\label{fig:fig7}}
\end{figure*}

\clearpage

\bibliographystyle{aasjournalv7}
\bibliography{mybib}

\end{document}